\documentclass[11pt,a4paper,leqno]{article}

\usepackage{latexsym}
\usepackage{amssymb, amsmath}

\newif\ifpdf
\ifx\pdfoutput\undefined
   \pdffalse
\else
   \pdfoutput=1
   \pdftrue
\fi

\ifpdf
   \usepackage[pdftex,
               pdftitle={On the coherence of Expected Shortfall},
               pdfsubject={How to create a coherent version of Expected Shortfall?},
               pdfkeywords={Expected Shortfall; Risk measure; worst conditional expectation; tail conditional expectation;
               value-at-risk (VaR); conditional value-at-risk (CVaR); tail mean; coherence; quantile;
               sub-additivity},
               pdfauthor={Carlo Acerbi, Dirk Tasche},
               pdfstartview=FitBH,
               pdfview=FitBH,
               colorlinks=true,
               citecolor=blue,
               urlcolor=magenta]{hyperref}
\else
   \usepackage{hyperref}
\fi

\newtheorem{theorem}{Theorem}[section]

\newtheorem{example}[theorem]{Example}
\newtheorem{proposition}[theorem]{Proposition}
\newtheorem{definition}[theorem]{Definition}
\newtheorem{corollary}[theorem]{Corollary}

\renewcommand{\theequation}{\arabic{section}.\arabic{equation}}

\title{On the coherence of Expected Shortfall}
\author{
Carlo Acerbi\thanks{Abaxbank, Corso Monforte 34, 20122 Milano, Italy;
E-mail: carlo.acerbi@abaxbank.com}
\quad
Dirk Tasche\thanks{Deutsche Bundesbank, Postfach 10 06 02, 60006 Frankfurt,
Germany; E-mail: tasche@ma.tum.de\newline
The contents of this paper do not necessarily reflect opinions shared by Deutsche Bundesbank.}
}

\date{April 19, 2002}

\begin{document}
\maketitle
\begin{abstract}
Expected Shortfall (ES) in several variants has been proposed as remedy for
the deficiencies of Value-at-Risk (VaR) which in general is not a coherent risk measure.
In fact, most definitions of ES lead to the same results when applied to
continuous loss distributions. Differences may appear when the underlying
loss distributions have discontinuities. In this case even the coherence
property of ES can get lost unless one took care of the details in its
definition. We compare some of the definitions of Expected Shortfall,
pointing out that there is one which is robust in the sense 
of yielding a coherent risk measure regardless of the underlying
distributions. Moreover, this Expected Shortfall can be estimated
effectively even in cases where the usual estimators for VaR fail.


\end{abstract}

\section{Introduction}

Value-at-Risk (VaR) as a risk measure is heavily criticized for not being
sub-additive (see \cite{Em00} for an overview of the
criticism). 
This means that the risk of a portfolio can be larger than
the sum of the stand-alone risks of its components when measured by VaR
(cf. \cite{ADEH97}, \cite{ADEH99}, \cite{RK99}, or \cite{ANS}). Hence, managing risk
by VaR may fail to stimulate diversification. Moreover, VaR does not take into
account the severity of an incurred damage event.

As a response to these deficiencies the notion of \emph{coherent} risk measures was
introduced in \cite{ADEH97}, \cite{ADEH99}, and \cite{Delb98}. An important example for
a risk measure of this kind is the \emph{worst conditional expectation} (WCE) 
(cf. Definition
5.2 in \cite{ADEH99}). This notion is closely related to the \emph{tail conditional expectation} (TCE)
from Definition 5.1 in \cite{ADEH99}, but in general does not coincide with it (see section \ref{sec:ineq} below).
Unfortunately, 
a somewhat misleading formulation in \cite{ADEH97} suggests this coincidence to be true.
Meanwhile, several authors (e.g. \cite{Ur00}, \cite{Pf00}, or \cite{ANS}) proposed modifications to TCE, 
this way increasing confusion
since the relation of these modifications to TCE and WCE remained obscure to a
certain degree.

The identification of TCE and WCE is to a certain degree 
a temptation though the authors of \cite{ADEH99} actually did their best to warn the reader. 
WCE is in fact coherent but very useful only in a theoretical setting since it requires the knowledge of the whole 
underlying probability space
while TCE lends itself naturally to practical applications but it is not coherent (see Example \ref{ex:ineq} below).
The goal to construct a risk measure which is  both coherent and easy to
compute and to estimate   
was however achieved in \cite{ANS}. The definition of \emph{Expected Shortfall} (ES) at a specified level $\alpha$ in \cite{ANS} 
(Definition \ref{def:6} below)
is the literal mathematical transcription of the concept ``average loss in the worst $100 \alpha$\% cases''. 
We rely on this definition of Expected Shortfall in the present paper, despite the fact that 
in the literature this term was already used sometimes in another meaning.

With the paper at hand we strive primarily for  making transparent the relations between the
notions developed in \cite{Delb98}, \cite{ADEH99},  \cite{Pf00}, and \cite{ANS}.
We present four characterizations of Expected shortfall: as integral of all the quantiles
below the corresponding level (eq.~(\ref{eq:3})), as limit in a tail
strong law of large numbers (Proposition \ref{pr:limit}), as minimum of a
certain functional introduced in \cite{Pf00} (Corollary \ref{co:1} below), and
as
maximum of $\mathrm{WCE}$s when the underlying probability space varies
(Corollary \ref{co:3}).
This way, we will show that the $\mathrm{ES}$ definition in \cite{ANS} is complementary 
and even in some aspects superior to the other notions. Moreover, in a certain sense any
law invariant coherent risk measure has a representation with $\mathrm{ES}$ as the main
building block (see \cite{Kusuoka}).

Some hints on the organization of the paper:\\ 
In section~\ref{sec:1.2} we give precise mathematical definitions to
the five notions to be discussed. These are WCE, TCE, CVaR \emph{(conditional
value-at-risk)}, ES, and its negative, the so-called
\emph{$\alpha$-tail mean} (TM).
Section~\ref{sec:properties} presents useful properties of $\alpha$-tail mean and ES, namely the
integral representation (\ref{eq:3}), continuity and
monotonicity in the level $\alpha$ as well as coherence for ES.
In section \ref{sec:2} we show first that $\alpha$-tail mean arises naturally as limit of the average of the $100 \alpha$\% worst
cases in a sample. Then we point out that in fact ES and CVaR are two different names for the same 
object. Section~\ref{sec:ineq} is devoted to inequalities and examples clarifying
the relations between ES, TCE, and WCE. In Section~\ref{sec:wce} we deal with the question how to state
a general representation of ES in terms of WCE. Section~\ref{sec:con} concludes the paper.

\section{Basic definitions}
\label{sec:1.2}
\setcounter{equation}{0}

We have to arrange a minimum set of definitions to be consistent with
the notions used in \cite{Delb98},  \cite{Pf00}, and \cite{ANS}.
Fix for this section some real-valued random variable $X$ on a probability space $(\Omega, \mathcal{A},
\mathrm{P})$. $X$ is considered the random profit or loss of some asset or portfolio. For the purpose
of this paper, we are mainly interested in losses, i.e. low values of $X$. 
By $\mathrm{E}[\ldots]$ we will denote expectation with respect to $\mathrm{P}$.
Fix also some confidence level $\alpha \in (0,1)$.
We will often make use of the \emph{indicator function}
\begin{equation}\label{eq:5}
\mathbf{1}_A(a)\ =\ \mathbf{1}_A\ =\ \left\{
  \begin{array}{c@{\,,\ }l}
1 & a \in A\\
0 & a \not\in A\,.
  \end{array}
\right.
\end{equation}

\begin{definition}[Quantiles]
  \label{def:1}\ \\
$x_{(\alpha)} = q_\alpha(X) = \inf \{ x\in \mathbb{R}:\ \mathrm{P}[X \le x] \ge \alpha \}$ is the
\emph{lower $\alpha$-quantile} of $X$,\\
$x^{(\alpha)} = q^\alpha(X) = \inf \{ x\in \mathbb{R}:\ \mathrm{P}[X \le x] > \alpha \}$ is the
\emph{upper $\alpha$-quantile} of $X$.\\
We use the $x$-notation if the dependence on $X$ is evident, otherwise the $q$-notion. \hfill $\Box$
\end{definition}

Note that $x^{(\alpha)} = \sup \{ x\in \mathbb{R}:\ \mathrm{P}[X \le x] \le \alpha \}$.
From $\{ x\in \mathbb{R}:\ \mathrm{P}[X \le x] > \alpha \} \subset  \{ x\in \mathbb{R}:\ \mathrm{P}[X \le x] \ge \alpha \}$
it is clear that $x_{(\alpha)} \le x^{(\alpha)}$. Moreover, it is easy to see that
\begin{equation}
x_{(\alpha)} \ =\ x^{(\alpha)}\quad \mbox{if and only if}\quad \mathrm{P}[X \le x] \ = \ \alpha\quad
\mbox{for at most one}\ x\,,
  \label{eq:bd1}
\end{equation}
and in case  $x_{(\alpha)} < x^{(\alpha)}$
\begin{equation}
 \{ x \in \mathbb{R}:\  \alpha = \mathrm{P}[X \le x]\} \ = \
\left\{
  \begin{array}{c@{\,,\quad}l}
{[} x_{(\alpha)},\, x^{(\alpha)}) & \mathrm{P}[ X = x^{(\alpha)}] > 0\\
{[} x_{(\alpha)},\, x^{(\alpha)}] & \mathrm{P}[ X = x^{(\alpha)}] = 0\,.
  \end{array}
\right.
  \label{eq:bd2}
\end{equation}
(\ref{eq:bd1}) and (\ref{eq:bd2}) explain why it is difficult to say
that there is an obvious definition for \emph{value-at-risk (VaR)}.
We join here \cite{Delb98} taking as $\mathrm{VaR}^\alpha$ the smallest value such
that the probability of the absolute loss being at most this value
is at least $1-\alpha$. As this is not really comprehensible when said
with words here is the formal definition:

\begin{definition}[Value-at-risk]
  \label{def:2}\ \\
$\mathrm{VaR}^\alpha = \mathrm{VaR}^\alpha(X) = - x^{(\alpha)} = q_{1-\alpha}(-X)$ is the \emph{value-at-risk}
at level $\alpha$ of $X$. \hfill $\Box$ 
\end{definition}

The definition of \emph{tail conditional expectation (TCE)} given in
\cite{ADEH99}, 
Definition 5.1,  depends on the choice
of quantile taken  for VaR (and of some discount factor we neglect here for reasons of simplicity).
But as there is a choice for VaR there is also a choice for TCE. That is why we consider a lower and
an upper TCE.
Denote the positive part of a number $x$ by $x^+ = \left\{ 
  \begin{array}{c@{\,,\ }l}
x & x > 0\\
0 & x \le 0\,,
  \end{array}
\right.
$ and its negative part by $x^- = (- x)^+$.

\begin{definition}[Tail conditional expectations]\ \\
\label{def:3}
Assume $\mathrm{E}[X^-] < \infty$. Then
$\mathrm{TCE}_\alpha = \mathrm{TCE}_\alpha(X) = - \mathrm{E}[ X\,|\, X \le  x_{(\alpha)}]$ is the
\emph{lower tail conditional expectation} at level $\alpha$ of $X$.\\
$\mathrm{TCE}^\alpha = \mathrm{TCE}^\alpha(X) = - \mathrm{E}[ X\,|\, X \le  x^{(\alpha)}]$ is the
\emph{upper tail conditional expectation} at level $\alpha$ of $X$. \hfill $\Box$
\end{definition}

$\mathrm{TCE}^\alpha$ is (up to a discount factor) the tail conditional expectation from Definition 5.1
in \cite{ADEH99}. ``Lower'' and ``upper'' here corresponds to the quantiles used for the definitions,
but not to the proportion of the quantities. In fact,
\begin{equation}
\mathrm{TCE}_\alpha\quad \ge\quad \mathrm{TCE}^\alpha
  \label{eq:bd3}
\end{equation}
is obvious.

As Delbaen says in the proof of Theorem 6.10 in \cite{Delb98}, $\mathrm{TCE}^\alpha$ in general does
not define a sub-additive risk measure (see Example \ref{ex:ineq} below). For this
reason, in \cite{ADEH99}, Definition 5.2, the \emph{worst conditional expectation (WCE)} was introduced. Here
is the definition (up to a discount factor) in our terms:

\begin{definition}[Worst conditional expectation]
\label{def:4}\ \\
Assume $\mathrm{E}[X^-] < \infty$. Then
$\mathrm{WCE}_\alpha = \mathrm{WCE}_\alpha(X) = - \inf\{ \mathrm{E}[X\,|\,A]:\ A \in \mathcal{A}, 
\mathrm{P}[A] > \alpha\}$ is the \emph{worst conditional expectation} at level $\alpha$ of $X$.
\hfill $\Box$
\end{definition}

Observe that under the assumption $\mathrm{E}[X^-] < \infty$ the value of $\mathrm{WCE}_\alpha$ is always
finite since then $\lim_{t\to\infty} \mathrm{P}[X\le x^{(\alpha)}+t]=1$ implies that there is some
event $A = \{X\le x^{(\alpha)}+t\}$ with $\mathrm{P}[A] > \alpha$ and $\mathrm{E}[\,|X|\,\mathbf{1}_A] < \infty$.
We will see in section \ref{sec:ineq} that Definition \ref{def:4} has to be
treated with care nevertheless because the notion $\mathrm{WCE}_\alpha(X)$ hides the fact
that it depends not only on the distribution of $X$ but also on the structure
of
the underlying probability space.
From the definition it is clear that for any random variables $X$ and $Y$ on the same probability
space
$$
\mathrm{WCE}_\alpha(X + Y) \quad \le \quad \mathrm{WCE}_\alpha(X) + \mathrm{WCE}_\alpha(Y)\,,
$$
i.e. $\mathrm{WCE}$ is sub-additive. Moreover, Proposition 5.1 in \cite{ADEH99} says
$\mathrm{WCE}_\alpha \ge \mathrm{TCE}^\alpha$. Hence $\mathrm{WCE}_\alpha$ is a majorant
to $\mathrm{TCE}^\alpha \ge \mathrm{VaR}^\alpha$. It is in fact the smallest coherent
risk measure dominating $\mathrm{VaR}^\alpha$ and only depending on $X$ through its
distribution if the underlying probability space is ``rich'' enough 
(see Theorem 6.10 in \cite{Delb98} for details). 

This is a  nice result, but to a certain degree unsatisfactory since the infimum
does not seem too handy. This observation might have been the reason for introducing
the \emph{conditional value-at-risk (CVaR)} in \cite{Ur00} (see also
the references therein) and \cite{Pf00}. CVaR can be used as a base for very efficient optimization 
procedures. We quote here, up to the sign of the random variable and the 
corresponding change from $\alpha$ to $1 - \alpha$ (cf. Definition \ref{def:2}), equation (1.2) from \cite{Pf00}.

\begin{definition}[Conditional value-at-risk]
\label{def:5}\ \\
Assume $\mathrm{E}[X^-] < \infty$. Then
$\mathrm{CVaR}^\alpha = \mathrm{CVaR}^\alpha(X) =  \inf \left\{ \frac{\mathrm{E}[ (X - s)^- ]}{\alpha} - s \,:\ s \in \mathbb{R}
\right\}$ is the \emph{conditional value-at-risk} at level $\alpha$ of $X$.
\hfill $\Box$
\end{definition}
Note that  by Proposition \ref{pr:3} and (\ref{eq:3.3}), $\mathrm{CVaR}$ is well-defined.
But beware: Pflug states  in equation (1.3) of \cite{Pf00}  (translated to our setting, i.e.
$-X$ instead of $Y$ and $1 - \alpha$ instead of $\alpha$) the relation
$
\mathrm{CVaR}^\alpha(X) =  \mathrm{TCE}^\alpha(X)\,,
$
without any assumption.
Corollary \ref{co:ineq2} in connection with Corollary \ref{co:1} shows that this is only true 
if $\mathrm{P}[X < x_{(\alpha)}] = 0, \mathrm{P}[X = x^{(\alpha)}] = 0$ or $\mathrm{P}[X < x_{(\alpha)}] > 0, 
\mathrm{P}[X \le x_{(\alpha)}] = \alpha$ (in particular if the distribution of $X$ is continuous).

The last definition we need is that of \emph{$\alpha$-tail mean} from \cite{ANS}. In order to make
it comparable to the risk measures defined so far, we define it in two variants: the tail mean which is likely
to be negative but appears in a statistical context (cf. Proposition \ref{pr:limit} below), and the Expected Shortfall representing
potential loss as in most cases positive number.
The advantage of tail mean is the explicit representation allowing an easy proof of super-additivity (hence sub-additivity for its
negative)
independent of the distributions of the underlying random variables (cf. the theorem in the
appendix of \cite{ANS}). We will see below (Corollary \ref{co:1}) that the Expected Shortfall is in fact identical with
$\mathrm{CVaR}$ and enjoys properties  as coherence and continuity and
monotonicity in the 
confidence level (section \ref{sec:properties}). Moreover, it is in a specific
sense the largest possible value $\mathrm{WCE}$ can take (Corollary \ref{co:3}).

\begin{definition}[Tail mean and Expected Shortfall]
\label{def:6}\ \\
Assume $\mathrm{E}[X^-] < \infty$. Then\\
$\bar{x}_{(\alpha)} = \mathrm{TM}_\alpha(X) =  \alpha^{-1} \big( \mathrm{E}[ X \,\mathbf{1}_{\{X \le x_{(\alpha)}\}}]
+ x_{(\alpha)}\,(\alpha - \mathrm{P}[X \le x_{(\alpha)}])\Big)$ is the \emph{$\alpha$-tail mean} at level $\alpha$ of
$X$.\\
$\mathrm{ES}_\alpha = \mathrm{ES}_\alpha(X) = -\bar{x}_{(\alpha)}$ is the \emph{Expected Shortfall (ES)} at level $\alpha$ of
$X$. \hfill $\Box$
\end{definition}
Note that by Corollary \ref{co:1}  $\alpha$-tail mean  and $\mathrm{ES}_\alpha$ only depend on the distribution of $X$ and the level
$\alpha$ but not on a particular definition of quantile.


\section{Useful properties of tail mean and Expected Shortfall}
\label{sec:properties}
\setcounter{equation}{0}

The most important property of ES (Definition \ref{def:6}) might be its coherence.

\begin{proposition}[Coherence of ES]
\label{pr:coh}
Let $\alpha \in (0,1)$ be fixed. Consider a set $V$ of 
real-valued random variables on some probability space $(\Omega, \mathcal{A}, \mathrm{P})$
such that $\mathrm{E}[X^-] < \infty$ for all $X \in V$. Then $\rho: V \to \mathbb{R}$ with
$\rho(X) = \mathrm{ES}_\alpha(X)$ for $X\in V$
is a coherent risk measure in the sense of Definition 2.1 in \cite{Delb98}, i.e. it is
\begin{enumerate}
\item monotonous: $X \in V,\ X \ge 0 \quad \Rightarrow \quad \rho(X) \le 0$,
\item sub-additive: $X, Y, X+Y \in V\quad \Rightarrow \quad \rho(X+Y) \le
\rho(X) + \rho(Y)$,
\item positively homogeneous: $X \in V,\ h > 0, h\,X\in V \quad \Rightarrow \quad \rho(h\,X)
= h\,\rho(X)$, and
\item translation invariant: $X \in V, \ a \in \mathbb{R} \quad \Rightarrow \quad \rho(X + a) =
\rho(X) - a$.
\end{enumerate}  
\end{proposition}
\textbf{Proof.} See Proposition \ref{pr:A} in the Appendix for an elementary proof of (ii). To check (i),
(iii) and (iv) is an easy exercise (cf. also Proposition \ref{pr:integral}). \hfill $\Box$

In the financial industry there is a growing necessity to deal with random
variables with discontinuous distributions. Examples are  portfolios of
not-traded loans (purely discrete distributions) or portfolios 
containing derivatives (mixtures of continuous and discrete distributions).
One problem with tail risk measures like $\mathrm{VaR}$, $\mathrm{TCE}$, and $\mathrm{WCE}$,
when applied to discontinuous distributions,
may be their sensitivity to small changes in the confidence level $\alpha$.
In other words, they are not in general continuous with respect to the
confidence level $\alpha$ (see Example \ref{ex:ineq}).

In contrast, $\mathrm{ES}_\alpha$ is continuous with respect to $\alpha$. Hence,
regardless of the underlying distributions, one can be sure that the risk measured
by $\mathrm{ES}_\alpha$ will not change dramatically when there is a switch
in the confidence level by -- say -- some base points.
We are going to derive this insensitivity property in Corollary \ref{pr:cont}
below as a consequence of an alternative representation of tail mean. This integral representation
(Proposition \ref{pr:integral})
-- which was already given in \cite{BLS00} for the case of continuous distributions --
might be of interest on its own.
Another -- almost self-evident --  important property of $\mathrm{ES}_\alpha$
is its monotonicity in $\alpha$. The smaller the level $\alpha$ the greater is the
risk. We show this formally in Proposition
\ref{pr:non-decreasing}. 

\begin{proposition}\label{pr:integral}
If $X$ is a real-valued random variable on a probability space 
$(\Omega, \mathcal{A}, \mathrm{P})$ with $\mathrm{E}[X^-] < \infty$ and 
$\alpha \in (0,1)$ is fixed, then
$$
\bar{x}_{(\alpha)} \quad = \quad \alpha^{-1} \int_0^\alpha x_{(u)}\,d\,u\,,
$$
with $\bar{x}_{(\alpha)}$ and $x_{(u)}$ as in Definitions \ref{def:1} and 
\ref{def:6}, respectively. 
\end{proposition}
\textbf{Proof.} By switching to another probability space if necessary, we can 
assume that there is a real random variable $U$ on $(\Omega, \mathcal{A}, \mathrm{P})$ that
is uniformly distributed on $(0,1)$, i.e. $\mathrm{P}[U \le u] = u$, $u \in (0,1)$. 
It is well-known that then the random
variable $Z = x_{(U)}$ has the same distribution as $X$.

Since $u \mapsto x_{(u)}$ is non-decreasing we have
\begin{eqnarray}
  \label{eq:1}
\{ U \le \alpha\} & \subset & \{ Z \le x_{(\alpha)}\}\quad \mbox{and}\\
\label{eq:2}
\{ U >  \alpha\} \cap \{ Z \le x_{(\alpha)}\} & \subset &
\{ Z = x_{(\alpha)}\}\,.
\end{eqnarray}
By (\ref{eq:1}) and (\ref{eq:2}) we obtain
\begin{eqnarray*}
 \int_0^\alpha x_{(u)}\,d\,u & = & \mathrm{E}[Z\,\mathbf{1}_{\{U \le \alpha\}}]\\
& = & \mathrm{E}[Z\,\mathbf{1}_{\{Z \le x_{(\alpha)}\}}] 
- \mathrm{E}[Z\,\mathbf{1}_{\{ U >  \alpha\} \cap\{Z \le x_{(\alpha)}\}}] \\
& = & \mathrm{E}[X\,\mathbf{1}_{\{X \le x_{(\alpha)}\}}] + 
x_{(\alpha)}\,\Big( \alpha - \mathrm{P}[X \le x_{(\alpha)}]\Big).
\end{eqnarray*}
Dividing by $\alpha$ now yields the assertion. \hfill $\Box$

Note that by definition of Expected Shortfall, Proposition \ref{pr:integral}
implies the representation
\begin{equation}
  \label{eq:3}
\mathrm{ES}_\alpha(X) \quad = \quad - \,\alpha^{-1} \int_0^\alpha q_u(X)\,d\,u\,.
\end{equation}
Eq.~(\ref{eq:3}) shows that $\mathrm{ES}$ is the coherent risk measure used in \cite{Kusuoka}
as main building block for the representation of law invariant coherent risk measures.

\begin{corollary} \label{pr:cont} 
If $X$ is a real-valued random variable with $\mathrm{E}[X^-] < \infty$, then the mappings
$\alpha \mapsto \bar{x}_\alpha$ and $\alpha \mapsto \mathrm{ES}_\alpha$ are continuous on
$(0,1)$.
\end{corollary}
\textbf{Proof.} Immediate from Proposition \ref{pr:integral} and (\ref{eq:3}).
\hfill $\Box$

For some of the results below and in particular the subsequent proposition on
monotonicity 
of the tail mean and $\mathrm{ES}$, a further
representation for $\overline{x}_{(\alpha)}$ is useful (cf. Appendix in
\cite{ANS}). Let 
for $x \in \mathbb{R}$
\begin{equation}
\label{eq:repfunc}
{\bf 1}_{\{X\leq x\}}^{(\alpha)}\quad  =\quad
\left\{
\begin{array}{c@{\,,\ }l}
{\bf 1}_{\{X\leq x\}} & \mbox{if}\quad \mathrm{P}[X=x] = 0\\[1ex]
{\bf 1}_{\{X\leq x\}} 
+
\frac{\alpha -\mathrm{P}[X\le x] }{\mathrm{P}[X=x]} 
\,{\bf 1}_{\{X = x\}} & \mbox{if}\quad \mathrm{P}[X=x] > 0\,.
\end{array}
\right.
\end{equation}
Then a short calculation shows
\begin{eqnarray}
{\bf 1}_{\{X\leq x_{(\alpha)}\}}^{(\alpha)} & \in & [0,1]\,,
\label{in01}\\
\label{E=alpha}
\mathrm{E}\left[{\bf 1}_{\{X\leq x_{(\alpha)}\}}^{(\alpha)}\right]& =& \alpha\,,\quad \mbox{and}\\
 \label{repr}
\alpha^{-1} \mathrm{E}\left[ X \,{\bf 1}_{\{X\leq x_{(\alpha)}\}}^{(\alpha)}\right] & = & \overline{x}_{(\alpha)}\,.
\end{eqnarray}
 
\begin{proposition} \label{pr:non-decreasing} 
If $X$ is a real-valued random variable with $\mathrm{E}[X^-] < \infty$, then for any $\alpha \in (0,1)$ and any $\epsilon > 0$
with $\alpha + \epsilon < 1$ we have the following inequalities:
\begin{eqnarray*}
        \overline{x}_{(\alpha+\epsilon)} & \geq &
        \overline{x}_{(\alpha)} \qquad\mbox{and}\\
\mathrm{ES}_{\alpha+\epsilon}(X) & \leq & \mathrm{ES}_{\alpha}(X)\,.
\end{eqnarray*}
\end{proposition}
\textbf{Proof.} We adopt  the representation (\ref{repr}). This yields
\begin{eqnarray*}
\overline{x}_{(\alpha+\epsilon)} -      \overline{x}_{(\alpha)} &=&  
\mathrm{E}
\left[ 
X \,
\left(
(\alpha+\epsilon)^{-1}
{\bf 1}_{\{X\leq x_{(\alpha+\epsilon)}\}}^{(\alpha+\epsilon)} -
\alpha^{-1}
{\bf 1}_{\{X\leq x_{(\alpha)}\}}^{(\alpha)} 
\right)
\right] \\[1ex]
&=& 
(\alpha (\alpha+\epsilon))^{-1} 
\mathrm{E}
\left[ 
X \,
\left(
\alpha \,
{\bf 1}_{\{X\leq x_{(\alpha+\epsilon)}\}}^{(\alpha+\epsilon)} -
(\alpha+\epsilon) \,
{\bf 1}_{\{X\leq x_{(\alpha)}\}}^{(\alpha)} 
\right)
\right] \\[1ex]
&\geq& 
(\alpha (\alpha+\epsilon))^{-1} 
\mathrm{E}
\left[ 
x_{(\alpha)} \,
\left(
\alpha\,
{\bf 1}_{\{X\leq x_{(\alpha+\epsilon)}\}}^{(\alpha+\epsilon)} -
(\alpha+\epsilon) \,
{\bf 1}_{\{X\leq x_{(\alpha)}\}}^{(\alpha)} 
\right)
\right] \\[1ex]
&=& 
\frac{x_{(\alpha)}}{\alpha (\alpha+\epsilon)} 
\left(
\alpha\, \mathrm{E}
\left[ {\bf 1}_{\{X\leq x_{(\alpha+\epsilon)}\}}^{(\alpha+\epsilon)}
\right]
- (\alpha+\epsilon) \, 
\mathrm{E}
\left[ {\bf 1}_{\{X\leq x_{(\alpha)}\}}^{(\alpha)}
\right]
\right) \\[1ex]
&=& 
\frac{x_{(\alpha)}}{\alpha (\alpha+\epsilon)} 
\left(
\alpha\, (\alpha+\epsilon)- (\alpha+\epsilon)\,\alpha  
\right) \\
&=& 0\,.
\end{eqnarray*}
The inequality is due to the fact that by (\ref{in01})

\hfill
$\displaystyle
\alpha \,
{\bf 1}_{\{X\leq x_{(\alpha+\epsilon)}\}}^{(\alpha+\epsilon)} -
(\alpha+\epsilon) \,
{\bf 1}_{\{X\leq x_{(\alpha)}\}}^{(\alpha)} 
\  
\left\{
\begin{array}{c@{\,,\quad}l}
\leq\ 0 & \mbox{if}\ X<x_{(\alpha)} \\ 
\geq\ 0 & \mbox{if}\ X>x_{(\alpha)}\,.
\end{array}
\right.
$
\hfill $\Box$


\section{Motivation for tail mean and Expected Shortfall}
\label{sec:2}
\setcounter{equation}{0}

Assume that we want to estimate the lower $\alpha$-quantile $x_{(\alpha)}$ of some random
variable $X$. Let some sample $(X_1, \ldots, X_n)$, drawn from independent copies of $X$,
be given. Denote by
$X_{1:n} \le \ldots \le X_{n:n}$ the components of the ordered $n$-tuple $(X_1, \ldots, X_n)$.
Denote by $\lfloor x\rfloor$ the integer part of  the number $x\in\mathbb{R}$, hence
$$
\lfloor x\rfloor \quad =\quad  \max\{n\in\mathbb{Z}:\ n \le x\}\,.
$$
Then the order statistic $X_{\lfloor n \alpha\rfloor:n}$ 
appears as natural estimator for $x_{(\alpha)}$.
Nevertheless, it is well known that in case of a non-unique quantile (i.e.
$x_{(\alpha)} < x^{(\alpha)}$) the quantity $X_{\lfloor n \alpha\rfloor:n}$ 
does not converge to $x_{(\alpha)}$. This follows for instance from Theorem~1 in \cite{FT66} which says
that
\begin{eqnarray*}
1\ = \  \mathrm{P}[X_{\lfloor n \alpha\rfloor:n} \le x_{(\alpha)}\ \mbox{infinitely often}] & 
= &   \mathrm{P}[X_{\lfloor n \alpha\rfloor:n} \ge x^{(\alpha)}\ \mbox{infinitely often}]\,.
\end{eqnarray*}

Surprisingly, we get a well-determined limit when we replace the single order statistic
by an average over the left tail of the sample. Recall the definition (\ref{eq:5}) of an indicator function.

\begin{proposition} \label{pr:limit}
Let $\alpha \in (0,1)$ be fixed, $X$ a real random variable with $\mathrm{E}[X^-] < \infty$ and $(X_1, X_2, \ldots)$ an
independent sequence of random variables with the same distribution as $X$.  Then
with probability 1
\begin{equation}
\lim_{n\to\infty} \frac{\sum\limits_{i=1}^{\lfloor n \alpha \rfloor} X_{i:n}}{\lfloor n \alpha \rfloor}
\quad  = \quad  \bar{x}_{(\alpha)}\,.
  \label{eq:9}
\end{equation}
If $X$ is integrable, then the convergence in (\ref{eq:9}) holds  in $\mathrm{L}_1$, too.
\end{proposition}
\textbf{Proof.} Due to Proposition \ref{pr:integral}, the ``with probability 1'' part of 
Proposition~\ref{pr:limit} 
is essentially a special case of Theorem~3.1 in \cite{vZwet80} with
$0 = t_0 < \alpha = t_1 < t_2 = 1$, $J(t) = \mathbf{1}_{(0, \alpha]}(t)$, 
$J_N(t) = \mathbf{1}_{(0, \frac{\lfloor N \alpha\rfloor + 1}N]}(t)$, $g(t) = 
F^{-1}(t)$,
and $p_1 = p_2 = \infty$. Concerning the $\mathrm{L}_1$-convergence note that
$$
\bigl| \sum\limits_{i=1}^{\lfloor n \alpha \rfloor} X_{i:n}\bigr| \quad \le \quad
\sum\limits_{i=1}^n |X_i|.
$$
By the strong law of large numbers $n^{-1} \sum_{i=1}^n |X_i|$ converges in $\mathrm{L}_1$.
This implies uniform integrability for $n^{-1} \sum_{i=1}^n |X_i|$ and for
$n^{-1} \bigl| \sum_{i=1}^{\lfloor n \alpha \rfloor} X_{i:n}\bigr|$. Together with the already
proven almost sure convergence this implies the assertion.
\hfill $\Box$

To see how a direct proof  of the almost sure convergence in 
Proposition~\ref{pr:limit} would work consider the following heuristic
computation. Observe first that
\begin{eqnarray}
 \frac{\sum\limits_{i=1}^{\lfloor n\alpha\rfloor} X_{i:n} }{\lfloor n\alpha\rfloor}
&=& 
\frac{1}{\lfloor n\alpha\rfloor}
\left(
\sum_{i=1}^{n} X_{i:n}\, {\bf 1}_{\{X_{i:n}\leq X_{\lfloor n\alpha\rfloor:n}\}} +
\sum_{i=1}^{n}
X_{i:n}
\Big(
{\bf 1}_{\{1, \ldots, \lfloor n\alpha\rfloor\}}(i)-
{\bf 1}_{\{X_{i:n}\leq X_{\lfloor n\alpha\rfloor:n}\}}
\Big)
\right)\nonumber \\[2ex]
&=& 
\frac{1}{\lfloor n\alpha\rfloor}
\left(
\sum_{i=1}^{n} X_{i}\,{\bf 1}_{\{X_{i}\leq X_{\lfloor n\alpha\rfloor:n}\}} +
X_{\lfloor n\alpha\rfloor:n}
\sum_{i=1}^{n}
\Big(
{\bf 1}_{\{1, \ldots, \lfloor n\alpha\rfloor\}}(i)-
{\bf 1}_{\{X_{i:n}\leq X_{\lfloor n\alpha\rfloor:n}\}}
\Big)
\right)\nonumber \\[2ex]
&=& 
\frac{1}{\lfloor n\alpha\rfloor}
\left(
\sum_{i=1}^{n} X_{i}\, {\bf 1}_{\{X_{i}\leq X_{\lfloor n\alpha\rfloor:n}\}} +
X_{\lfloor n\alpha\rfloor:n}
\Big(
\lfloor n\alpha\rfloor -
\sum_{i=1}^{n}
{\bf 1}_{\{X_{i}\leq X_{\lfloor n\alpha\rfloor:n}\}}
\Big)
\right).\label{eq:decomp} 
\end{eqnarray}
If we now had 
\begin{equation}
\lim_{n\to\infty} X_{\lfloor n\alpha\rfloor:n} \quad = \quad x_{(\alpha)}\,,
  \label{eq:lim}
\end{equation}
with probability 1,
in connection with $\lim_{n\to\infty} n / \lfloor n\alpha\rfloor = 1/\alpha$ it would be
plausible to obtain (\ref{eq:9}). Unfortunately (\ref{eq:lim}) is not true in general,
but only
\begin{equation}
\liminf_{n\to\infty} X_{\lfloor n\alpha\rfloor:n} \quad = \quad x_{(\alpha)} \qquad\mbox{and}\qquad
\limsup_{n\to\infty} X_{\lfloor n\alpha\rfloor:n}\quad = \quad x^{(\alpha)}\,.
  \label{eq:liminf}
\end{equation}
Nevertheless the proof could be completed on the base of (\ref{eq:decomp}) by using
(\ref{eq:liminf}) together with the Glivenko-Cantelli theorem and Corollary \ref{co:1} below.

Proposition \ref{pr:limit} validates the interpretation given to $\alpha$--tail mean in \cite{ANS} as 
{\em mean of the worst $100 \alpha$\% cases}. This concept, which seems very natural from an 
insurance or risk management point of view, has so far  
appeared in the literature by different kinds of {\em conditional expectation beyond $\mathrm{VaR}$} 
which is a different concept for discrete distributions. ``Tail Conditional expectation'', 
``worst conditional expectation'', ``conditional value at risk'' all bear also in their name 
the fact that they are conditional expected values of the random variable $X$ (note that concerning 
$\mathrm{CVaR}$, by Corollary \ref{co:1} below this is a misinterpretation). For $\mathrm{TCE}_\alpha$, 
for instance, the natural estimator is not given by the one analyzed in (\ref{eq:9}) or its negative, but rather by
\begin{equation}
-\, \frac{\sum_{i=1}^n \, X_i \, {\bf 1}_{\{X_i \leq X_{\lfloor n\alpha \rfloor:n}\}}}{\sum_{i=1}^n \,  
{\bf 1}_{\{X_i \leq X_{\lfloor n\alpha \rfloor:n}\}}}
\end{equation}
which however has problems of convergence in case $x_{(\alpha)} < x^{(\alpha)}$.

This is the reason why we avoid the term "conditional" in our definition of $\alpha$--tail mean. In fact, it is not very hard
 to see (cf. Example \ref{ex:ineq} below) that $\alpha$--tail mean 
{\em does not} admit  a general representation in terms of a conditional expectation of $X$
given some event $A \in \sigma(X)$ (i.e. some event only depending on $X$). Hence it is not possible to give a definition of the type
\begin{equation}
\bar{x}_{(\alpha)}\quad = \quad
\mathrm{E}[ X | A]  \hspace{10mm} \mbox{for some}\ A \in  \sigma(X) \, ,
\end{equation}
unless the event $A$ is chosen  in a  $\sigma$-algebra $\mathcal{A} \supset \sigma(X)$ 
on an artificial new probability space (see Corollary \ref{co:2} below).


In order to make visible the coincidence of $\mathrm{CVaR}$ and tail mean, the following proposition collects some facts on quantiles 
which are well-known in probability theory (cf. Exercise~3 in ch.~1 of
\cite{Fer67} or  
Problem 25.9 of \cite{Hin72} for the here cited version):

\begin{proposition}\label{pr:3}
Let $X$ be a real  integrable random variable
on some probability space $(\Omega, {\cal A}, \mathrm{P})$. Fix $\alpha \in (0,1)$
and define the function $H_\alpha: \mathbb{R} \to [0,\infty)$ by 
\begin{equation}
H_\alpha(s) \quad = \quad \alpha\,\mathrm{E}[ (X - s)^+ ] + (1-\alpha)\,\mathrm{E}[ (X - s)^- ]\,.
  \label{eq:3.1}
\end{equation}
Then the function $H_\alpha$ is convex (and hence continuous) with
$\displaystyle
\lim_{|s|\to \infty} H_\alpha(s) = \infty$. The set $M_\alpha$ of minimizers to
$H_\alpha$ is a compact interval, namely \\
\begin{minipage}[b]{13cm}
\begin{eqnarray}
M_\alpha & = &  [ x_{(\alpha)},\, x^{(\alpha)}]
\nonumber  \\
\label{eq:3.2}
& = & \{ s \in \mathbb{R}:\ \mathrm{P}[X < s] \le \alpha \le \mathrm{P}[X \le s]\}\,.
\end{eqnarray}
\end{minipage}
\hfill
\parbox[t]{1cm}{\hfill $\Box$}
\end{proposition}

Note the following equivalent representations for $H_\alpha$:
  \begin{eqnarray}
H_\alpha(s) & = &  \alpha\,\mathrm{E}[X] + \alpha\,\left( \frac{\mathrm{E}[ (X - s)^- ]}{\alpha} - s \right) 
    \label{eq:3.3}\\
& = &  \alpha\,\mathrm{E}[X] - \alpha\,\left( \frac{\mathrm{E}[ X\,\mathbf{1}_{\{X \le s\}} ]}{\alpha} + s\,\frac{\alpha - 
\mathrm{P}[X \le s]}{\alpha} \right)\,. 
    \label{eq:3.4}
  \end{eqnarray}

From Definitions  \ref{def:5} and \ref{def:6} for $\mathrm{CVaR}$ and  $\mathrm{ES}$, respectively, and by (\ref{eq:3.4}), in connection
with Proposition \ref{pr:3}, we obtain the following corollary to  the proposition.
\begin{corollary}
  \label{co:1}
Let $X$ be a real  integrable random variable
on some probability space $(\Omega, {\cal A}, \mathrm{P})$ and $\alpha \in (0,1)$ be fixed.
Then\\
\begin{minipage}[b]{14cm}
\begin{eqnarray}
\mathrm{ES}_\alpha(X) & = & 
\mathrm{CVaR}^\alpha(X)\nonumber
\\
& = & -\,\alpha^{-1} \left( \mathrm{E}[ X\,\mathbf{1}_{\{X \le s\}} ] + s\,(\alpha - 
\mathrm{P}[X \le s])\right),\quad s \in  [ x_{(\alpha)},\, x^{(\alpha)}]\,.
\label{eq:const}
\end{eqnarray}
\end{minipage}
\hfill
\parbox[t]{1cm}{\hfill $\Box$}
\end{corollary}
A further representation of $\mathrm{ES}$ or $\mathrm{CVaR}$, respectively, as expectation
of a suitably modified tail distribution is given
in the recent research report \cite{RU01} (cf. Def.~3 therein).

In Definitions \ref{def:5} and \ref{def:6} only $\mathrm{E}[X^-] < \infty$ is required for $X$.
Indeed, this integrability condition would suffice to guarantee (\ref{eq:const}). We formulated
Corollary \ref{co:1} with full integrability of $X$ because we wanted to rely on Proposition \ref{pr:3}
for the proof.

Note that by a simple calculation one can show that (\ref{eq:const}) is equivalent to
\begin{eqnarray}
\mathrm{ES}_\alpha(X) 
& = & - \,\alpha^{-1} \left( \mathrm{E}[ X\,\mathbf{1}_{\{X < s\}} ] + s\,(\alpha - 
\mathrm{P}[X < s])\right),\quad s \in  [ x_{(\alpha)},\, x^{(\alpha)}]\,. \label{eq:3.7b}
\end{eqnarray}
By (\ref{eq:3.7b}) we see that $\mathrm{ES}$ coincides with the coherent risk measure considered
in Example~4 of \cite{Delb01} (already mentioned in Example 4.2 of \cite{Delb98}).


\section{Inequalities and counter-examples}
\label{sec:ineq}
\setcounter{equation}{0}

In this section we compare the Expected Shortfall with the risk measures
$\mathrm{TCE}$ and $\mathrm{WCE}$ defined in section \ref{sec:1.2}. 
Moreover, we present an example showing that $\mathrm{VaR}$ and $\mathrm{TCE}$ 
are not sub-additive in general. By the same example we show that
there is not a clear relationship between $\mathrm{WCE}$ and lower $\mathrm{TCE}$.
We start with a result in the spirit of the Neyman-Pearson lemma.

\begin{proposition}
  \label{pr:ineq}
Let $\alpha \in (0,1)$ be fixed and $X$ be a real-valued random variable on some
probability space $(\Omega, \mathcal{A}, \mathrm{P})$. Suppose that
there is some function $f:\mathbb{R}\to\mathbb{R}$ such that
$\mathrm{E}[(f\circ X)^-]<\infty$, $f(x) \le f(x_{(\alpha)})$ for $x <
x_{(\alpha)}$, and $f(x) \ge f(x_{(\alpha)})$ for $x > x_{(\alpha)}$. 
Let $A\in \mathcal{A}$ be an event with $\mathrm{P}[A]\ge \alpha$ and
$\mathrm{E}[\,|f\circ X|\,\mathbf{1}_A] < \infty$. Then
\begin{enumerate}
\item $\mathrm{TM}_\alpha(f\circ X) \ \le \ \mathrm{E}[f\circ X\,|\,A]\,,$
\item  $\mathrm{TM}_\alpha(f\circ X) \ = \ \mathrm{E}[ f\circ X\,|\,A]$\quad if\
$\mathrm{P}[A\cap \{X > x_{(\alpha)}\}] = 0$\ and
  \begin{eqnarray}
    \label{eq:pr1}
&&\mathrm{P}[X < x_{(\alpha)}] = 0 \quad \mbox{or} \\
&&\mathrm{P}[X < x_{(\alpha)}] > 0,\ \mathrm{P}[\Omega \backslash A \cap \{X <
x_{(\alpha)}\}] = 0, \ \mbox{and}\quad \mathrm{P}[A]=\alpha\,,
\label{eq:pr1a}
  \end{eqnarray}
\item if $f(x) < f(x_{(\alpha)})$ for $x <
x_{(\alpha)}$ and $f(x) > f(x_{(\alpha)})$ for $x > x_{(\alpha)}$, then
$\mathrm{TM}_\alpha(f\circ X) \ = \ \mathrm{E}[ f\circ X\,|\,A]$  implies
$\mathrm{P}[A\cap \{X > x_{(\alpha)}\}] = 0$ and either 
(\ref{eq:pr1}) or (\ref{eq:pr1a}).
\end{enumerate}
\end{proposition}
\textbf{Proof.} Note that by assumption 
$$
\{X \le x_{(\alpha)}\} \subset \{f\circ X \le f(x_{(\alpha)})\}\quad
\mbox{and}\quad \{X < x_{(\alpha)}\} \supset \{f\circ X < f(x_{(\alpha)})\}\,.
$$
Hence we see from (\ref{eq:3.2}) that 
$$\mathrm{P}[f\circ X \le f(x_{(\alpha)})] \ge \alpha\quad \mbox{and}\quad
\mathrm{P}[f\circ X < f(x_{(\alpha)})] \le \alpha
$$
and therefore 
\begin{equation}\label{eq:pr2}
q_{\alpha}(f\circ X)\ \le \ f(x_{(\alpha)})\ \le \ q^{\alpha}(f\circ X)\,.
\end{equation}
Moreover, the assumption implies 
\begin{equation}\label{eq:pr2a}
\{f\circ X \le f(x_{(\alpha)})\}\backslash \{X \le x_{(\alpha)}\}
\quad \subset \quad \{f\circ X = f(x_{(\alpha)})\}\,.
\end{equation}
By Corollary \ref{co:1}, (\ref{eq:pr2}), (\ref{eq:pr2a}), 
(\ref{E=alpha}), and (\ref{repr}), we can calculate similarly 
to the proof of Proposition \ref{pr:non-decreasing} 
\begin{eqnarray}
\mathrm{E}[ f\circ X| A ] - \mathrm{TM}_\alpha(f\circ X)   &=&
\mathrm{E}
\left[ 
f\circ X \,
\left(
P[A]^{-1}
{\bf 1}_{A} -
\alpha^{-1}
{\bf 1}_{\{f\circ X\leq f(x_{(\alpha)})\}}^{(\alpha)} 
\right)
\right]\nonumber\\[1ex]
&=&
\mathrm{E}
\left[ 
f\circ X \,
\left(
P[A]^{-1}
{\bf 1}_{A} -
\alpha^{-1}
{\bf 1}_{\{ X\leq x_{(\alpha)}\}}^{(\alpha)} 
\right)
\right]\nonumber\\[1ex]
& = & (\alpha \,P[A])^{-1} \bigg( f(x_{(\alpha)})
\mathrm{E}
\left[ 
\alpha \,
{\bf 1}_{A} - P[A]
 \,
{\bf 1}_{\{X\leq x_{(\alpha)}\}}^{(\alpha)} 
\right] \nonumber\\[1ex]
& & \quad +\
\mathrm{E}
\left[ 
(f\circ X - f(x_{(\alpha)})) \,
\left(
\alpha \,
{\bf 1}_{A} - P[A]
 \,
{\bf 1}_{\{X\leq x_{(\alpha)}\}}^{(\alpha)} 
\right)
\right]\bigg)\nonumber\\[1ex]
& = & (\alpha \,P[A])^{-1} 
\mathrm{E}
\left[
(f\circ X - f(x_{(\alpha)})) \,
\left( 
\alpha \,
{\bf 1}_{A} - P[A]
 \,
{\bf 1}_{\{X\leq x_{(\alpha)}\}}^{(\alpha)} 
\right)
\right] \nonumber\\[1ex]
&\geq& 
0\,.
  \label{eq:le2}
\end{eqnarray}
Here, we obtain inequality (\ref{eq:le2}) from the assumption on $f$ since
\begin{equation}
\alpha \,
{\bf 1}_A - 
\mathrm{P}[A] \,
{\bf 1}_{\{X\leq x_{(\alpha)}\}}^{(\alpha)} 
\  
\left\{
\begin{array}{c@{\,,\quad}l}
\leq\ 0 & \mbox{if}\ X<x_{(\alpha)} \\ 
\geq\ 0 & \mbox{if}\ X>x_{(\alpha)}\,.
\end{array}
\right.
  \label{eq:le3}
\end{equation}
This proves (i). The sufficiency and necessity respectively of the conditions
in (ii) and (iii) for equality in (\ref{eq:le2}) are easily obtained by
careful inspection
of (\ref{eq:le2}).
\hfill $\Box$

Note that the condition
\begin{equation}\label{eq:ineq10a} 
\mathrm{P}[A\cap \{X > x_{(\alpha)}\}] = 0 \quad \mbox{and}\quad
 \mathrm{P}[\Omega \backslash A \cap \{X <
x_{(\alpha)}\}] = 0\,,
\end{equation}
appearing in (ii) and (iii) of Proposition \ref{pr:ineq}, means up to
set differences of probability 0 that
\begin{equation}
\{ X < x_{(\alpha)}\} \quad \subset \quad A \quad \subset \quad \{ X \le x_{(\alpha)}\}\,.
\label{eq:ineq10}
\end{equation}
In particular, (\ref{eq:ineq10a}) is implied by (\ref{eq:ineq10}).

The proof of Proposition \ref{pr:ineq} is the hardest work in this section.
Equipped with its result we are in a position to derive without effort
a couple of conclusions pointing out the relations between TCE, WCE and ES.
Recall $\mathrm{ES}_\alpha = - \mathrm{TM}_\alpha$.

\begin{corollary}
  \label{co:ineq1}
Let $\alpha \in (0,1)$ and $X$ a real-valued random variable on some
probability space $(\Omega, \mathcal{A}, \mathrm{P})$ with $\mathrm{E}[X^-] < \infty$.
Then
\begin{eqnarray}
\mathrm{TCE}^\alpha(X) & \le &  \mathrm{TCE}_\alpha(X) \quad \le \quad
\mathrm{ES}_\alpha(X)\,, \quad \mbox{and}
  \label{eq:ineq11}\\
\mathrm{TCE}^\alpha(X) & \le & \mathrm{WCE}_\alpha(X) \quad \le \quad
\mathrm{ES}_\alpha(X)\,.
\label{eq:ineq11a}
\end{eqnarray}
\end{corollary}
\textbf{Proof.} The first inequality in (\ref{eq:ineq11}) is obvious (formally it
follows from Lemma 5.1 in \cite{T00}). The second
follows from Proposition \ref{pr:ineq} (i) by 
setting $f(x) = x$, $A = \{X \le x_{(\alpha)}\}$, and observing 
$\mathrm{P}[X \le x_{(\alpha)}]\ge \alpha$.
The first inequality in (\ref{eq:ineq11a}) was proven in Proposition 5.1 of \cite{ADEH99}.
The second follows again from Proposition \ref{pr:ineq} (i) since all the
events
in the definition of $\mathrm{WCE}$ have probabilities $> \alpha$.
\hfill $\Box$ 

The following corollary to Proposition \ref{pr:ineq} presents in particular in (i)
a first sufficient condition for $\mathrm{WCE}$ and $\mathrm{ES}$ to coincide,
namely continuity of the distribution of $X$.

\begin{corollary}
  \label{co:ineq2}
Let $\alpha$ and $X$ be as in Corollary \ref{co:ineq1}. Then
\begin{enumerate}
%
\item $\mathrm{P}[X \le x^{(\alpha)}]  =
  \alpha,\ \mathrm{P}[X < x_{(\alpha)}] > 0$ or
$\mathrm{P}[X \le x^{(\alpha)},\, X \not= x_{(\alpha)}] = 0$   
 \quad
if and only if
\begin{equation}\label{eq:4=}
\mathrm{ES}_\alpha(X) \ = \ \mathrm{WCE}_\alpha(X) \ = \ \mathrm{TCE}_\alpha(X)\ = \ \mathrm{TCE}^\alpha(X)\,.
\end{equation}
In particular,  (\ref{eq:4=}) holds if the distribution of $X$ is continuous,
i.e.
$\mathrm{P}[X = x] = 0$ for all $x\in\mathbb{R}$.
\item $\mathrm{P}[X \le x_{(\alpha)}] = \alpha$ or $\mathrm{P}[X <
  x_{(\alpha)}] = 0$ \quad if and only if
\quad $\mathrm{ES}_\alpha(X) \ = \ \mathrm{TCE}_\alpha(X)$.
\end{enumerate}
\end{corollary}
\textbf{Proof.} 
Concerning (i)  apply Proposition \ref{pr:ineq} (ii) and (iii) with $A = \{ X
\le x^{(\alpha)}\}$
and Corollary \ref{co:ineq1}.
In order to obtain (ii) apply  Proposition \ref{pr:ineq} (ii) and (iii) with $A = \{ X \le x_{(\alpha)}\}$.
\hfill $\Box$

Corollary \ref{co:ineq1} leaves open the relation between $\mathrm{TCE}_\alpha(X)$ and $\mathrm{WCE}_\alpha(X)$.
The implication
\begin{equation}
\mathrm{P}[X \le x_{(\alpha)}]\ >\ \alpha\quad \Rightarrow\quad \mathrm{TCE}_\alpha(X) \le \mathrm{WCE}_\alpha(X)
  \label{eq:ineq13}
\end{equation}
is obvious. Corollary \ref{co:ineq2} (ii) shows that
\begin{equation}
\mathrm{P}[X \le x_{(\alpha)}]\ =\ \alpha\quad \Rightarrow\quad \mathrm{TCE}_\alpha(X) \ge \mathrm{WCE}_\alpha(X)\,.
  \label{eq:ineq13a}
\end{equation}

The following example shows that all the inequalities between $\mathrm{TCE}$, $\mathrm{WCE}$, and $\mathrm{ES}$
in (\ref{eq:ineq11}), (\ref{eq:ineq11a}), (\ref{eq:ineq13}), and (\ref{eq:ineq13a}) can be strict.
Moreover, it shows that none of the quantities $- q_{\alpha}$, $\mathrm{VaR}^{\alpha}$, $\mathrm{TCE}_\alpha$, or
$\mathrm{TCE}^\alpha$ defines a sub-additive risk measure in general.

\begin{example}
  \label{ex:ineq} \ \rm\\
Consider the probability space $(\Omega, \mathcal{A}, \mathrm{P})$ with
$\Omega = \{ \omega_1,\omega_2,\omega_3  \}$, $\mathcal{A}$ the set of all subsets of $\Omega$
and
$\mathrm{P}$ specified by 
$$
\mathrm{P}[\{\omega_1\}]=\mathrm{P}[\{\omega_2\}]= p, \quad \mathrm{P}[\{\omega_3\}]= 1-2\,p\,, 
$$
and choose $0<p<\frac{1}{3}$. Fix some positive number $N$ and
let $X_i$, $i=1,2$, be two random variables defined on $(\Omega, \mathcal{A}, \mathrm{P})$ with values
$$
X_i(\omega_j)\quad =\quad\left\{
\begin{array}{c@{\,,\quad}l}
- N & \mbox{if}\ i = j\\
0 & \mbox{otherwise.}
\end{array}
\right.
$$
Choose $\alpha$ such that $0 < \alpha<2\,p$. Then it is straightforward to
obtain Table \ref{tab} with the values of the risk measures interesting to us.

\begin{table}[h]
\centering
\begin{tabular}{|c|cc|cc|cc|}\hline
\multicolumn{1}{|c|}{}  & \multicolumn{2}{c|}{\bf{$p<\alpha<2p$}}& \multicolumn{2}{c|}{\bf{$p=\alpha$}} &
\multicolumn{2}{c|}{\bf{$p > \alpha$}} \\ \hline 
 \multicolumn{1}{|c|}{Risk Measure}        & \multicolumn{1}{c}{$X_{1,2}$}& \multicolumn{1}{c|}{$X_1+X_2$}
        & \multicolumn{1}{c}{$X_{1,2}$}& \multicolumn{1}{c|}{$X_1+X_2$} & \multicolumn{1}{c}{$X_{1,2}$}& \multicolumn{1}{c|}{$X_1+X_2$} \\ \hline 
$- q_{\alpha}$ & $0$  & $N$& $N$      & $N$ & $N$ & $N$           \\
$\mathrm{VaR}^{\alpha}$ & $0$  & $N$& $0$      & $N$ & $N$ & $N$           \\
$\mathrm{TCE}^{\alpha}$ & $N p$        & $N$& $N p$      & $N$  & $N$ & $N$                  \\
$\mathrm{TCE}_{\alpha}$ & $N p$        & $N$& $N$      & $N$   & $N$ & $N$                 \\
$\mathrm{WCE}_{\alpha}$ &  $N/2$       & $N$   & $N/2$   & $N$  & $N$ & $N$          \\
$\mathrm{ES}_{\alpha}$ &  $N p/\alpha$ & $N$   & $N$   & $N$   & $N$ & $N$         \\
 \hline 
\end{tabular}
\caption{Values of risk measures for Example \ref{ex:ineq}.}
\label{tab}
\end{table}

In case $p <\alpha<2\,p$ we see from Table \ref{tab} that
\begin{eqnarray*}
- q_\alpha{(X_1)} -  q_\alpha{(X_2)} & < &   - q_\alpha{(X_1 + X_2)}\\
\mathrm{VaR}^\alpha{(X_1)} +  \mathrm{VaR}^\alpha{(X_2)} & < &   \mathrm{VaR}^\alpha{(X_1 + X_2)}\\
\mathrm{TCE}_\alpha{(X_1)} +  \mathrm{TCE}_\alpha{(X_2)} & < &   \mathrm{TCE}_\alpha{(X_1 + X_2)}\\
\mathrm{TCE}^\alpha{(X_1)} +  \mathrm{TCE}^\alpha{(X_2)} & < &
\mathrm{TCE}^\alpha{(X_1 + X_2)}\,.
\end{eqnarray*}
These inequalities show that 
none of the notions  
$- q_{\alpha}$, $\mathrm{VaR}^{\alpha}$, $\mathrm{TCE}_\alpha$, or
$\mathrm{TCE}^\alpha$ can be used to define a sub-additive risk measure.
In case $p <\alpha<2\,p$ we have also
\begin{eqnarray}
\mathrm{TCE}_\alpha{(X_1)} & < &\mathrm{ES}_\alpha(X_1)\nonumber\\
\mathrm{TCE}^\alpha{(X_1)} \ = \ \mathrm{TCE}_\alpha{(X_1)} & < &\mathrm{WCE}_\alpha{(X_1)}\nonumber\\ 
\mathrm{WCE}_\alpha{(X_1)} & < &\mathrm{ES}_\alpha(X_1)\,. \label{eq:ineq14}
\end{eqnarray}
Hence the second inequalities in (\ref{eq:ineq11}), (\ref{eq:ineq11a}), and
(\ref{eq:ineq13})
may be strict, as can be the first inequality in (\ref{eq:ineq11a}).
In case $p =\alpha$ we have from Table \ref{tab} that 
\begin{eqnarray*}
\mathrm{TCE}^\alpha{(X_1)} & < &  \mathrm{TCE}_\alpha{(X_1)}\quad \mbox{and}\\
\mathrm{TCE}_\alpha{(X_1)} & > &  \mathrm{WCE}_\alpha{(X_1)}\,.
\end{eqnarray*}
Thus, also the first inequality in (\ref{eq:ineq11}) and the inequality in
(\ref{eq:ineq13a})
can be strict. In particular, we see that there is not any clear relationship
between
$\mathrm{TCE}_\alpha$ and $\mathrm{WCE}$.
Beside the inequalities, from the comparison with the results in the region $p>\alpha$, we get an example for the fact
that all the measures but ES may have discontinuities in $\alpha$. 
Moreover, in case $p < \alpha$ we have a stronger version of (\ref{eq:ineq14}), namely
$$
- \inf\{ \mathrm{E}[X_1\,|\,A]:\ A\in \mathcal{A}, \mathrm{P}[A]\ge \alpha\} \quad < \quad \mathrm{ES}_\alpha(X_1)\,,
$$
which shows that even if one replaces ``$>$'' by ``$\ge$'' in Definition \ref{def:4}, strict inequality may appear
in the relation between $\mathrm{WCE}$ and $\mathrm{ES}$. \hfill $\Box$
\end{example}

We finally observe that Example  \ref{ex:ineq} is not so academic as it may
seem at first glance since the 
$X_i$'s may be figured out as two risky bonds of nominal $N$ with non--overlapping default states $\omega_i$ of probability $p$.


\section{Representing ES in terms of WCE}
\label{sec:wce}
\setcounter{equation}{0}

By Example \ref{ex:ineq} we know that $\mathrm{WCE}$ and $\mathrm{ES}$ may differ in general.
Nevertheless, we are going to show in the last part of the paper that this phenomenon can
only occur when the underlying probability space is too ``small'' in the sense of not allowing
a suitable representation of the random variable under consideration as function of a continuous
random variable.
Moreover, as long as only finitely many random variables are under consideration it is always
possible to switch to a ``larger'' probability space in order to make
$\mathrm{WCE}$ and $\mathrm{ES}$ coincide. Finally, we state a general
representation
of $\mathrm{ES}$ in terms of related $\mathrm{WCE}$s.

\begin{proposition}\label{pr:wce} 
Let $X$ and $Y$ be a real-valued random variables  on a probability 
space $(\Omega, {\cal A}, \mathrm{P})$ such that  $\mathrm{E}[Y^-] < \infty$. Fix some $\alpha \in (0,1)$.
Assume that $Y$ is given by $Y = f \circ X$ where $f$ satisfies
$f(x) \le f(x_{(\alpha)})$ for $x <
x_{(\alpha)}$, and $f(x) \ge f(x_{(\alpha)})$ for $x > x_{(\alpha)}$.
\begin{enumerate}
\item If $\mathrm{P}[X \le x_{(\alpha)}] = \alpha$ then
$$
\mathrm{ES}_\alpha(Y) \quad= \quad - \inf_{A \in {\cal A},\, \mathrm{P}[A] \ge \alpha} \mathrm{E}[ Y\,|\, A]\,.
$$
\item If the distribution function of $X$ is continuous then also
$$
\mathrm{ES}_\alpha(Y) \quad= \quad \mathrm{WCE}_\alpha(Y)\,.
$$
\end{enumerate}
\end{proposition}
%
%
\textbf{Proof.} Concerning (i), by Proposition \ref{pr:ineq} (i) we only have to show
\begin{equation}
\mathrm{TM}_\alpha(Y) \quad = \quad \mathrm{E}[ Y\,|\, X \le x_{(\alpha)}]\,.
  \label{eq:ineq15}
\end{equation}
With the choice $A = \{X \le x_{(\alpha)}\}$ this follows from Proposition \ref{pr:ineq} (ii).

Concerning Proposition \ref{pr:wce} (ii), by (\ref{eq:ineq15}), we have to show
that there is a sequence $(A_n)_{n\in\mathbb{N}}$ in $\cal A$ with
$\mathrm{P}[ A_n ] > \alpha$ for all $n \in \mathbb{N}$ such that 
$$
\lim_{n\to \infty} \mathrm{E}[ Y\,|\,A_n] = \mathrm{E}[ Y\,|\,X \le  x_{(\alpha)}]\,.
$$
By  continuity of the distribution of $X$ and integrability of 
$Y^-$ we obtain such a sequence with the definition
$
A_n  =  \{ X \le x^{(\alpha)} + 1/n \}\,.
$
\hfill $\Box$ 

\begin{corollary}
  \label{co:2}
Let $(X_1, \ldots, X_d)$ be an $\mathbb{R}^d$-valued random vector on a
probability space $(\Omega, \mathcal{A}, \mathrm{P})$ such that
$\mathrm{E}[X_i^-] < \infty, i = 1, \ldots, d$. Fix $\alpha \in (0,1)$. Then there is a
random vector $(X'_1, \ldots, X'_d)$ on some probability space
$(\Omega', \mathcal{A'}, \mathrm{P'})$ with the following two
properties:
\begin{enumerate}
\item  The distributions of $(X_1, \ldots, X_d)$ and $(X'_1, \ldots, X'_d)$ are
equal, i.e. 
$$
\mathrm{P}[X_1 \le x_1, \ldots, X_d \le x_d ]\ =\
\mathrm{P'}[X'_1 \le x_1, \ldots, X'_d \le x_d ]\quad\mbox{for all}\
(x_1, \ldots, x_d) \in \mathbb{R^d}\,.
$$
\item Worst conditional expectation and Expected Shortfall coincide
for all $i = 1, \ldots, d$, i.e.\\
$\mathrm{WCE}_\alpha(X'_i) \ = \ \mathrm{ES}_\alpha(X'_i)\,,\ 
i=1,\ldots, d\,.$ \hfill $\Box$
\end{enumerate}
\end{corollary}
\textbf{Proof.} By Sklar's theorem (cf. Theorem 2.10.9 in \cite{Nel99}) we get the
existence of a random vector $(U_1, \ldots, U_d)$ where each $U_i$ is
uniformly distributed on $(0,1)$ such that (i) holds with
$X'_i = q_{U_i}(X_i), i = 1,\ldots, d$. Since $q_\alpha$ is non-decreasing
in $\alpha$ the assertion now  follows from Proposition \ref{pr:wce}.
\hfill $\Box$

Corollary \ref{co:2} yields another proof for the sub-additivity of
Expected Shortfall: in order to prove $\mathrm{ES}_\alpha(X) + \mathrm{ES}_\alpha(Y)
\ge\mathrm{ES}_\alpha(X+Y)$ apply the corollary to the underlying
random vector $(X, Y, X+Y)$. 

As a final consequence of Corollary \ref{co:ineq1} and Corollary \ref{co:2} we note:
\begin{corollary}
  \label{co:3}
Let $X$ be a real-valued random variable on some probability space 
$(\Omega, \mathcal{A}, \mathrm{P})$ with $\mathrm{E}[X^-] < \infty$.
Fix $\alpha \in (0,1)$. Then
\begin{eqnarray*}
\mathrm{ES}_\alpha(X) & = & \max \Big\{ \mathrm{WCE}_\alpha(X'): \ X' \
\mbox{random variable on} \ (\Omega', \mathcal{A}', \mathrm{P}')\ \mbox{with} \\
& & \hspace{4cm} \mathrm{P}'[X' \le x] = \mathrm{P}[X \le x]
\ \mbox{for all}\ x\in \mathbb{R}\Big\},
\end{eqnarray*}
where the maximum is taken over all random variables $X'$ on probability
spaces $(\Omega', \mathcal{A}', \mathrm{P}')$ such that the distributions
of $X$ and $X'$ are equal.\hfill $\Box$
\end{corollary}
Corollary \ref{co:3} shows that Expected Shortfall in the sense of Definition
\ref{def:6}
may be considered a robust version of worst conditional expectation
(Definition
\ref{def:4}), making
the latter insensitive to the underlying probability space.

\section{Conclusion}
\label{sec:con}
\setcounter{equation}{0}

In the paper at hand we have shown that simply taking a conditional 
expectation of losses beyond $\mathrm{VaR}$ can fail to yield a coherent
risk measure when there are discontinuities in the loss distributions.
Already existing definitions for some kind of expected shortfall, redressing
this drawback,
as those in \cite{ADEH99} or \cite{Pf00}, did not provide representations
suitable for efficient computation and estimation in the general case.
We have clarified the relations between these definitions and the explicit one
from \cite{ANS}, thereby pointing out that it is the definition which is
most appropriate for practical purposes.

\newpage

\appendix

\section{Appendix: Subadditivity of Expected Shortfall} \label{appendice1}

\renewcommand{\theequation}{{\rm{}A}.\arabic{equation}}

We give here for the sake of completeness the proof of subadditivity for expected shortfall which was originally given in Appendix A of \cite{ANS}. 

For the proof it is convenient to adopt the representation of eq. (\ref{repr}) for the Tail Mean and write the Expected Shortfall as
\begin{equation}
\mathrm{ES}_\alpha (Y) \: = \: - \:\frac{1}{\alpha}\: \mathrm{E}[ X\,\mathbf{1}_{\{X \le x_{(\alpha)}\}}^{(\alpha)}]
\end{equation}
with the function $\mathbf{1}_{\{X\leq s\}}^{(\alpha)}$ defined in (\ref{eq:repfunc}).

\begin{proposition}[Subadditivity of Expected Shortfall]%
\label{pr:A}%
Given two random variables $X$ and $Y$ with
$\mathrm{E}[X^-]<\infty$ and $\mathrm{E}[Y^-]<\infty$ the following inequality holds:
\begin{equation}
\mathrm{ES}_\alpha (X+Y)\ \leq\ \mathrm{ES}_\alpha (X) + \mathrm{ES}_\alpha (Y) 
\end{equation}
for any $\alpha \in (0,1]$
\end{proposition}
\textbf{Proof.} Defining $Z = X+Y$, we obtain by virtue of {(\ref{E=alpha})
\begin{align}
& \alpha\,%
\bigl( \mathrm{ES}_\alpha(X) + \mathrm{ES}_\alpha(Y)-\mathrm{ES}_\alpha(Z)\bigr) \: = \\\nonumber \\
&= 
{\bf E}\left[
Z\; \mathbf{1}_{\{Z \le z_{(\alpha)}\}}^{(\alpha)} - X\; \mathbf{1}_{\{X \le x_{(\alpha)}\}}^{(\alpha)} -Y\; \mathbf{1}_{\{Y \le y_{(\alpha)}\}}^{(\alpha)}
\right] \nonumber\\\nonumber\\
 & =  
{\bf E}\left[
X\;\left(  \mathbf{1}_{\{Z \le z_{(\alpha)}\}}^{(\alpha)} -  \mathbf{1}_{\{X \le x_{(\alpha)}\}}^{(\alpha)} \right)
+
Y\;\left(  \mathbf{1}_{\{Z \le z_{(\alpha)}\}}^{(\alpha)} -  \mathbf{1}_{\{Y \le y_{(\alpha)}\}}^{(\alpha)} \right)
\right] \nonumber\\\nonumber\\
&\geq  
x_{(\alpha)} \; {\bf E}\left[
 \mathbf{1}_{\{Z \le z_{(\alpha)}\}}^{(\alpha)} -  \mathbf{1}_{\{X \le x_{(\alpha)}\}}^{(\alpha)} \right]
+
y_{(\alpha)} \; {\bf E}\left[
 \mathbf{1}_{\{Z \le z_{(\alpha)}\}}^{(\alpha)} -  \mathbf{1}_{\{Y \le y_{(\alpha)}\}}^{(\alpha)} \right]
\nonumber\\\nonumber\\
& = x_{(\alpha)} \, (\alpha-\alpha) + y_{(\alpha)}\, (\alpha-\alpha) =0
\nonumber
\end{align}
which proves the thesis. In the inequality above we used the fact that
\begin{equation}
\left\{
\begin{array}{lll}
\displaystyle \mathbf{1}_{\{Z \le z_{(\alpha)}\}}^{(\alpha)} -  \mathbf{1}_{\{X \le x_{(\alpha)}\}}^{(\alpha)} \geq 0 & \hspace{5mm}\mbox{if} \hspace{5mm} & X>x_{(\alpha)} \\\\
\displaystyle \mathbf{1}_{\{Z \le z_{(\alpha)}\}}^{(\alpha)} -  \mathbf{1}_{\{X \le x_{(\alpha)}\}}^{(\alpha)} \leq 0 & \hspace{5mm}\mbox{if} \hspace{5mm} & X<x_{(\alpha)}
\end{array}
\right.
\end{equation}
which in turn is a consequence of (\ref{eq:repfunc}) and
(\ref{in01})
\hfill $\Box$

\newpage

\sloppy

\end{document}